# Silicon detectors for the next generation of high energy physics experiments: expected degradation

**I. Lazanu**

University of Bucharest, Faculty of Physics, POBox MG-11, Bucharest-Măgurele, Romania
e-mail: i_lazanu@yahoo.co.uk

**S. Lazanu**
National Institute for Materials Physics, POBox MG-7, Bucharest-Măgurele, Romania,
e-mail: lazanu@infim.ro

**Abstract**

There exists an enormous interest for the study of very high energy domain in particle physics, both theoretically and experimentally, in the aim to construct a general theory of the fundamental constituents of matter and of their interactions.
Until now, semiconductor detectors have widely been used in modern high energy physics experiments. They are elements of the high resolution vertex and tracking system, as well as of calorimeters.
The main motivation of this work is to discuss how to prepare some possible detectors – only silicon option being considered, for the new era of HEP challenges because the bulk displacement damage in the detector, consequence of irradiation, produces effects at the device level that limit their long time utilisation, increasing the leakage current and the depletion voltage, eventually up to breakdown, and thus affecting the lifetime of detector systems.
In this paper, physical phenomena that conduce to the degradation of the detector are discussed and effects are analysed at the device level (leakage current and effective carrier concentration) in the radiation environments expected in the next generation of hadron colliders after LHC, at the next lepton and gamma-gamma colliders, as well as in astroparticle experiments, in conditions of long time continuum irradiations, for different technological options. The predicted results permit a better decision to obtain devices with harder parameters to radiation.

## 1. New Physics Directions and Experimental Facilities

In the last fifty years physics has made striking advances in describing the intimate structure of matter and the forces that determine the architecture of the universe. The simple structure of elementary constituents and forces forms the theoretical framework called the "Standard Model", which has been able to predict with very good accuracy the values of many quantities that have been measured by experiments at modern accelerators. But this theory is incomplete because: gravity is not included, it requires as input a large number of parameters, some aspects, as for example electroweak symmetry breaking, the striking asymmetry between the massless photon and the massive W and Z, the difference by many orders of magnitude between the masses of fundamental constituents or the number of flavour families are not yet completely understood.

Experimental research in order to clarify these understood fundamental aspects is performed in the main areas: accelerator based particle physics, high energy astrophysics, astroparticle experiments and the study of the cosmology of the early universe. To obtain relevant answers to this quest higher energies are necessary, but this is not restricted to the highest energies, because at least observable traces of this new physics can be accessible also at low energy scales. So, new collider projects and studies of astroparticle physics represent two complementary links to the same scope.

Discussions on the search for new physics at the future collider facilities are presented in references [1] and [2]; a review of this thematic at the interface between particle and astroparticle physics was done by Pinfold [3].

The Large Hadron Collider (LHC) at CERN will be operational from 2007. But the physics results expected from this facility will not be able to clarify all the unknowns of the Standard Model. So, despite the technological difficulty, significant upgrades of the accelerator in energy and luminosity are already considered as Super-LHC and Very-LHC respectively. SLHC will represent an upgrade of the LHC collider while VLHC will be a new planned facility. The upgrade path will be defined by the results from the initial years of LHC operation, and now only possible scenarios can be done up to 240 TeV, the final energy of the project.
The Ultimate Large Hadron Collider (ULHC) [4] is imagined as accelerating protons at energies above $10^7$ TeV in a ring with a circumference L≈4×$10^4$ km around Earth and such a facility would permit to explore the "intermediate" energy scale, between the electroweek scale (1 TeV) and the scale of grand unification (around $10^{14}$ TeV).

Lepton colliders in the TeV scale would seem to be the tool of precision measurement of the new physics in this energy range. Requirements from physics are discussed in [5]. Unfortunately the performance of a multi - TeV $e^+e^-$ collider is severely limited by breamstrahlung and (for circular machine) by synchrotron radiation. Two solutions are possible to minimise these effects: to build electron linear colliders [6], [7] or muon colliders [8]. Both alternatives are considered in present paper.

The International Linear Collider (ILC) should be the next large scale project operating after LHC and with distinct physics goal. In the first stage a centre of mass energy of 500 GeV will be available, upgradable to about 1 TeV. The luminosity will be around $10^{34}$ $cm^{-2}s^{-1}$, corresponding to 200÷500 $fb^{-1}$/year. The concrete realisation depends on the physics needs. At the present time different variants are discussed.

The idea of high energy $\gamma-\gamma$ and $\gamma-e$ colliders with real photons on the base of a linear $e^{+/-}e^-$ collider is not new. In linear colliders each bunch is used only once. It makes possible to convert electrons to high energy photons and to obtain $\gamma-\gamma$ or $\gamma-e$ colliders with beams with approximately the same energy and luminosity as in the primary interaction processes. In these colliders, photon energy will be in the range from 100 GeV to several TeV, the number of high energy photons must be about $10^{10}$ per bunch with a transverse size at the interaction point (IP) close to $10^{-5}÷10^{-7}$ cm. Detailed analysis of this subject has been performed by Telnov [9] and Serbo [10].

In the last period arguments in favour of lepton-hadron (as eLHC) and photon-hadron colliders in the same TeV energy scale are discussed. A review of this proposal and of its advantages for crucial clarification of the dynamics of strong interaction could be found, for example, in Ref. [11] and in the references cited therein. These possible facilities are not considered here.

The main parameters of the future accelerator colliders, relevant for the present investigation, are presented in Table 1. For LHC upgrade, details could be found in references [12]] and [13]. .In Table 1, some characteristics of the background radiation in the interaction cavity where detectors work are also included. Details will be discussed in the next paragraph.

**Table 1.**
Collider parameters and characteristics of the background radiation for facilities in construction or planed

| Parameters | LHC | SLHC (stage 1) | SLHC (stage 2) | VLHC (stage 1) | VLHC (stage 2) | ULHC | ILC | μ+ μ- |
|---|---|---|---|---|---|---|---|---|
| $\sqrt{s}$ [TeV] | 14 | 14 | 28 | 40 | 175 (up to 240) | $2\times10^7$ | up to 1 | 4 |
| Circumference [km] | 26.7 | 26.7 | 26.7 | 233 | 233 | 40000 | | |
| Luminosity $\times 10^{34}$ [cm$^{-2}$s$^{-1}$] | 1 | 10 | 10 | 10 | 10 | ? 10 | 1.45 | 4.55 |
| Rep. frequency [Hz] | | | | | | | 120 | |
| Particle per bunch | | | | | | | $3.6\times10^{10}$ | |
| Bunch spacing [ns] | 25 | 12.5 | | 18.8 | 18.8 | | 1.4 | 186000 |
| $\sigma_{pp}$-total [mb] | ≈ 117 | | ≈ 136 | ≈ 143 | ≈ 166 | 770 | | |
| $\sigma_{pp/ee/\mu\mu}$-inel. [mb] | ≈ 80 | | ≈ 90 | ≈ 100 | ≈ 140 | ≈ 270 | $34\times10^{-6}$ | $54\times10^{-6}$ |
| Interactions | 20 | 100 | | | Up to 250 | | | |
| **Radiation background** | | | | | | | | |
| $\langle E_T \rangle$ charged part.[MeV] | 450 | 450 | 500 | 506 | 600 | 1820 | | |
| $\langle E \rangle$ components [MeV] | [1] γ: 30 e$^{+/-}$: 150 μ: 6600 h$^{+/-}$: 8100 n: 310 | | | | | | [2] γ: 8700 [3] | [4] γ: 1÷2.5 e$^{+/-}$: 80 μ: 130÷3630 h$^{+/-}$: 249 n: 0.2÷10 p: 30 π: 240 |

[1] Energy spectra of particles coming to the detector from LHC tunnel - from reference [14].
[2] Photons from beamstrahlung (the bending of the beam-particle trajectories leads to the emission of photons). 2.4 photons per e$^{+/-}$ particles are produced.
[3] $N_h=0.46$ is number of hadronic events per beam particle, $N_\perp=46.5$ is number of particle that have transverse momentum > 20 MeV/c and an angle with respect to beam axis of more than 150 mrad. From references [15] and [6]
[4] Mean energies of background particles in the tracking volume - results from Ref. [16]. Intervals represent the differences between GEANT and MARS simulations.

Another possibility to obtain information about the interactions of ultra high energy particles, complementary to colliders, is the exploitation of astroparticles or cosmic rays (CR). This supposes experiments with cosmic beams and terrestrial detectors, or complete experiments in space. We concentrate only to investigate effects to detectors in space. Galactic cosmic rays originating far outside our solar system are the most typical cosmic rays. They are the origin of the inner radiation belt, with their flux modulated by solar activity or as solar energetic particles. Protons are the most abundant charged particles in space. The CR proton spectrum, in the kinetic energy range up to hundred GeV, in the neighbourhood of Earth, as have been measured by the Alpha Magnetic Spectrometer during space shuttle flight STS-91 [17] for example, and their spectra was simulated by DESIRE Collaboration [18]. Extreme high energy CR particles, produced by different mechanisms, are not considered because their flux is irrelevant to produce any effect of degradation in detectors. It must be mentioned that the domain of energy between $10^5 \div 10^8$ TeV of the ultra high energy-CR, interpreted as the onset of extragalactic proton dominance in the data [19] is closed by the expected energy of protons at ULHC collider.

In this work, the expected behaviour of detectors - only silicon option being considered, for this new era of HEP challenges, is discussed. Due to the existence, long time, of intense radiation fields in the environments where detectors work, bulk displacement damage in silicon is produced, and consequently effects appear at the device level that limit their long time utilisation, increasing the leakage current and the depletion voltage sometimes up to breakdown, and thus affecting the lifetime of detector systems.

## 2. Radiation Background for Detectors

The overall detector performance in high energy experiments is dependent on the background particle rates in various detector components. The knowledge of mutual effects of the radiation environment produced by the accelerator and experiments represents one of the key issues in the development of detectors.
Despite different colliding particle types and machine parameters, there are many common features of the radiation background in detectors at these accelerators.
Particles originating from the interaction point and collision remnants are most often the major source of background in detectors at the hadron or lepton colliders. Beam loss in the vicinity of the interaction point is the second source of background.

In silicon detectors, radiation damage is essentially correlated with the flux of particles and their energies in background, because they contribute to the rate of generation of primary defects in the material. Also, high instantaneous particle fluxes complicate track reconstruction and cause increased trigger rates and affect detector occupancy. The existence long time of high radiation levels will produce radio activation of detector components as well as radiation environment in the experimental hall and in its surroundings.

In the LHC case, inside the tracker cavity, hadrons represent about 54% of all produced particles, and the pions are the most abundant, around 64% of all hadrons [20]. In the present work, the simulated charged hadron spectra at different positions inside the tracking cavity, for the concrete CMS future experiment, are used, from Reference [21]. The position (r=20 cm, z=0÷60 cm) which corresponds to the maximum in flux is considered. This particular choice does not affect the main conclusions, because similar radiation fields are predicted also by the ATLAS Collaboration. For LHC upgrade (SLHC and VLHC), the following hypotheses were considered: increase in luminosity and increase in the energy of the beam that, in agreement with Gianotti's evaluations [13], conduces to higher values of the mean energy of the spectra. For this case, we supplementary supposed that the shape of differential energetic spectra of pions and protons remains unmodified, and is only shifted to higher energy. The same scenario is considered also for the ULHC case.

At lepton colliders, the background arises from beam – beam effects, synchrotron radiation and muons. As we pointed out before, to create a $\gamma\gamma$ or a $\gamma e$ collider, their parameters must be comparable to those in $e^+e^-$ colliders. The main sources of background at photon colliders are hadrons, $e^+e^-$ pairs and photons. Hadrons are produced in $\gamma + \gamma \rightarrow hadrons$ processes. With a cross section of about 500 nb, for a typical luminosity of $10^{34}$ cm$^{-2}$s$^{-1}$ in an electron linear collider, the background rate is 0.5 events/bunch crossing. Electrons/positrons are produced as pairs or as low energy electrons after multiple Compton scattering (disrupted beams). If these low energy electrons can be removed from the IP, thus, the level of $e^+$ and $e^-$ in the vertex detector is approximately the same as for $e^+e^-$ collider [9], eventually with an additional contribution due the particles reflected by mirrors. The photons come from Compton scattering, beamstrahlung or from halo of X rays.

For concrete estimation, following Telnov's and Serbo's conclusions we consider that at the electron linear collider and photon collider the luminosity is the same as at LHC, but the hadronic background rate is up to 5 orders of magnitude lower. We also suppose that the spectra are similar.

The case of cosmic rays is different. If one considers that silicon detectors will be exposed directly in space, thus, in essence their degradation is produced by primary cosmic protons and supplemented by different contributions from other particles; if the detectors will be placed in a cavity, the contribution from background must also be considered.

## 3. Interactions of Particles in the Detector and the Kinetics of Defects

The nuclear interaction between the incident particle and the lattice nuclei of the detector produces bulk defects. After this interaction process, the recoil nucleus or nuclei are displaced from lattice positions into interstitials, in their initial positions remaining vacancies. Then, the primary knock-on nucleus, if its energy is large enough, could produce the displacement of a new nucleus, and the process continues as long as the energy of the colliding nucleus is higher than the threshold for atomic displacements. The physical quantity characterising the process is the concentration of primary defects produced per unit of fluence of the incident particles (CPD(E)).

The primary recoil is either a nucleus of the medium (produced in an elastic or quasielastic interaction process), or a nucleus with a lower mass and charge numbers - after inelastic interactions. As a consequence, for each medium a whole family of curves, $L(E_R)$, function of incident energy, i.e. energy spent into non-ionising processes versus recoil energy, could be obtained. These are the Lindhard curves and are used in the evaluation of the damage produced in materials. The concentration of the primary radiation induced defects per unit of particle fluence could be calculated using the explicit formula:

$$CPD(E) = \frac{N}{2E_d} \int \sum_i \left(\frac{d\sigma}{d\Omega}\right)_i L(E_{Ri}) d\Omega \qquad (1)$$

where $E$ is the kinetic energy of the incident particle, $N$ is the atomic density of the target element, $A$ is the atomic number of the silicon target, $E_d$ - the threshold energy for displacements (with values function of the defined symmetry plane), $E_{Ri}$ - the recoil energy of the residual nucleus produced in interaction $i$, $L(E_{Ri})$ - the Lindhard factor that describes the partition of the recoil energy between ionisation and displacements and $\left(d\sigma/d\Omega\right)_i$ - the differential cross section of the interaction between the incident particle and the nucleus of the lattice for the process or mechanism $(i)$, responsible in defect production. The energy dependences of CPD for different incident particles in silicon are presented in Figure 1

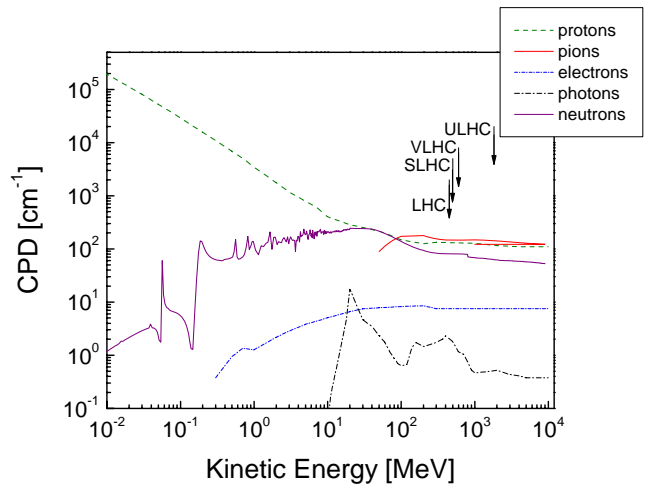

**Figure 1.**
Concentration of primary defects in silicon as function of the energy of primary particle

After this step, the physical quantity that is essential for degradation processes is the rate of generation of defects in the material of the detector, by irradiation, defined as:

$$G_{irrad} = \sum_{all\ particles} \int CPD(E)\Phi(E)dE \qquad (2)$$

where $\Phi(E)$ is the differential flux for the considered particles.

It must be mentioned that in $G_{irrad}$, the identity of the primary particles which initiated the bulk process is lost.

For hadron colliders considered here, the expected averaged energy, $<E_T>$, corresponds to values for CPD(E) in the vicinity of its asymptotic high energy limit:

$$CPD^{asymp}(E) \cong \frac{N}{2E_d} \sum_i L(E_{Ri})\sigma_i , \qquad (3)$$

as could be seen in Figure 1. In this region, the values of CPD are nearly equal for different hadrons.

In the case of $\mu^+\mu^-$ collider, the mean values of energies of electrons correspond to their asymptotic region in CPD. Also, in accord with Geer's simulations [16], at 5 cm from vertex there will be $1.69\times10^{-7}$ photons, $8.4\times10^{-11}$ electrons, $1.6\times10^{-9}$ neutrons, per act of muon - muon interaction, and $4.4\times10^{-10}$ ($\gamma$) and $1.5\times10^{-9}$ (n) in the tracker, at 50 cm. Other contributions are irrelevant.

In the last years, the CR proton spectrum, in the neighbourhood of Earth, has been measured by the AMS experiment during the space shuttle flight STS-91 and published in Ref. [17]. In reference [18], Ersmark and co-workers presented the belt proton spectra at $51.6^0$ and at 330, 380 and 430 km, simulated with GEANT4, for the International Space Station orbit in free space as well as the effects due the presence of the station. The spectra of secondary particles entering the Columbus cabin have also been simulated and were considered in the present analysis. Primary cosmic ray protons have energies in the range between some MeV up to $10^5$ GeV.

In Table 2, we present the calculated values for the rate of generation of primary defects in silicon, in the radiation fields corresponding to different future facilities.

**Table 2.**
Contributions of different particles at the rate of generation of defects
in silicon in the radiation environments of the new facilities / experiments.

| Source | Rate of generation of defects due to irradiation [pairs/cm$^3$/sec] | | | | | |
|---|---|---|---|---|---|---|
| | **Protons** | **Neutrons** | **Pions** | **Electrons** | **Photons** | **Total** |
| **New colliders** | | | | | | |
| LHC | 5.6x10$^7$ | | 6.2x10$^8$ | | | 6.8x10$^8$ |
| SLHC (stage 1) | 5.6x10$^8$ | | 6.2x10$^9$ | | | 6.8x10$^9$ |
| SLHC (stage 2) | 3.0x10$^8$ | | 5.1x10$^9$ | | | 5.4x10$^9$ |
| VLHC (stage 1) | 2.9x10$^8$ | | 5.0x10$^9$ | | | 5.3x10$^9$ |
| VLHC (stage 2) | 2.2x10$^8$ | | 3.5x10$^9$ | | | 3.7x10$^9$ |
| ULHC | 6.x10$^7$ | | 8.1x10$^8$ | | | 8.7x10$^8$ |
| ILC/gamma-gamma[1)] | | | | | | <10$^4$ |
| $\mu^+\mu^-$ | 1.6x10$^6$ | 3.3x10$^9$ | | 4.4x10$^3$ | 6.x10$^3$ | ≈3.3x10$^9$ |
| **Cosmic rays** | | | | | | |
| Belt protons (at 330km) | 1.x10$^5$ | | | | | 1.x10$^5$ |
| Belt protons (at 380km) | 1.9x10$^5$ | | | | | 1.9x10$^5$ |
| Belt protons (at 430km) | 3.4x10$^5$ | | | | | 3.4x10$^5$ |
| Belt and solar protons | 7.2x10$^5$ | | | | | 7.2x10$^5$ |
| Particle spectra in Columbus cabin | 68 | 220 | 30.7 | 3.8 | 3.2 | ≈326 |

[1)] Estimation based on Serbo's and Telnov's considerations.

The total rates of generation of defects include also thermal contributions. In all considered situations, a $20^0$ C temperature is supposed.

CPD is not proportional to the modifications of material parameters after irradiation, due to the subsequent interactions of vacancies and interstitials and with other defects and impurities (only phosphorus, carbon and oxygen are considered as pre-existents) in the silicon lattice.

The formation and time evolution of complex defects, associations of primary defects or of primary defects and impurities are studied in accord with the model developed in successive steps by the authors; see for example [22] and previous results cited therein.

The primary point defects considered here are silicon self interstitials and vacancies: "classic" vacancies and fourfold coordinated vacancy defects - $Si_{FFCD}$. This new defect has been predicted by Goedecker and co-workers [23] and its characteristics were indirectly determined by Lazanu and Lazanu [24].

So,

$$Si \xrightarrow{G} (V + Si_{FFCD}) + I . \qquad (4)$$

The rate $G$ is a sum of contributions from thermal generation and irradiation. The phenomena occurring in the material after primary defect production are considered in the frame of the theory of diffusion limited reactions. The complete list of reactions considered is also presented in Ref. [24].

The model predicts absolute values for microscopic quantities – concentrations of defects which could be used to evaluate device degradation, without free parameters.

Until now, systematic experimental studies to put in evidence detector degradation were realised by exposing these devices in radiation fields at different irradiation facilities by step by step irradiations/relaxations, up to fixed fluences. In fact, these experimental setups differ from realistic conditions where detectors are exposed for long times to continuous irradiation fields. These experimental results were utilised as a useful tools to test the validity of the theoretical model developed by the authors and to understand the mechanisms of damage and the results were presented in Reference [25]. In spite of the simplicity of the model, a remarkable agreement was obtained between model calculations and experimental data for leakage current and effective carrier concentration, considering different materials (FZ and DOFZ silicon), with resistivities in the range 1÷15 kΩcm, and irradiated with protons, pions, neutrons and electrons . In the case of CZ silicon technology the model is able to reproduce the situations corresponding to low fluences (beginning of irradiation), but supplementary processes must be considered, most probable associated with the presence of oxygen in the material.

## 4. Expected Behaviour of Silicon Detectors in Radiation Fields at the Next Generation of High Energy Physics Experiments

In the analysis of the behaviour of silicon detectors in the radiation fields considered as background in the next high energy experiments, two materials were considered: FZ and DOFZ, with characteristics listed in Table 3.

**Table 3.**
Silicon characteristics used in the calculations

| Material technology | [P] atoms/cm$^3$ | [O] atoms/cm$^3$ | [C] atoms/cm$^3$ |
|---|---|---|---|
| FZ | 8·10$^{11}$ | 10$^{15}$ | 9·10$^{15}$ |
| DOFZ | 8·10$^{11}$ | 4·10$^{17}$ | 9·10$^{15}$ |

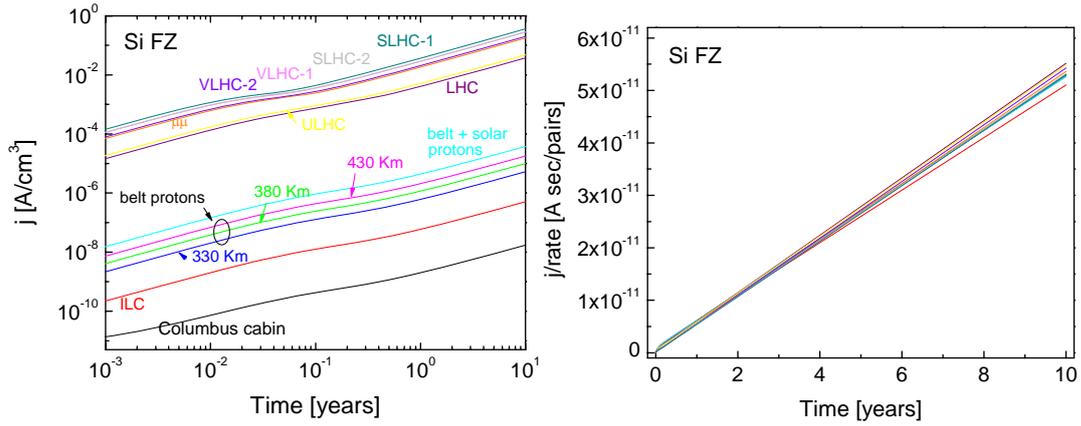

**Figure 2**
a) Time dependence of the leakage current in FZ silicon detectors in different radiation environments
b) Time dependence of the leakage current divided by the generation rate of primary defects in FZ silicon detectors

In **Figures 2a** and **3a** the calculated time dependence of the volume density of the leakage current in detectors produced from FZ and DOFZ Si, irradiated in the environments considered in Table 2 are presented. These materials show similar behaviours.

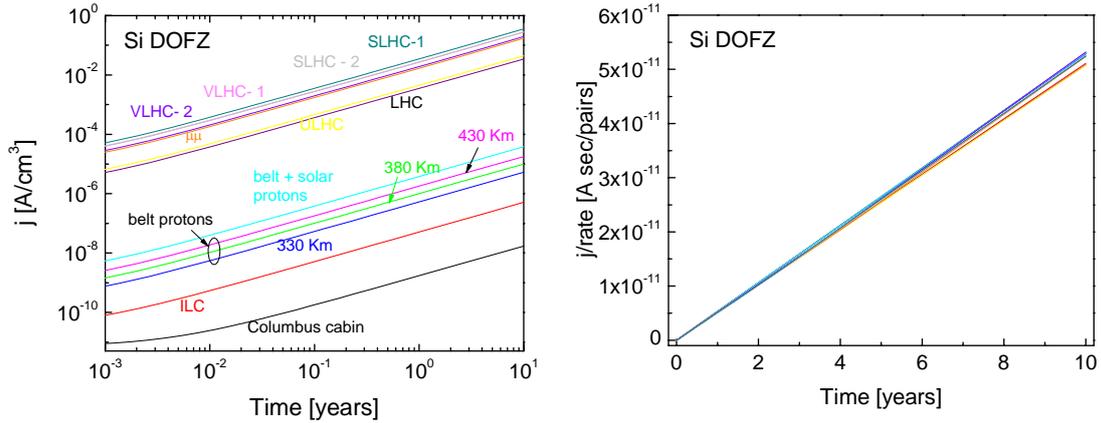

**Figure 3**
a) Time dependence of the leakage current in DOFZ silicon detectors in different radiation environments
b) Time dependence of the leakage current divided by the generation rate of primary defects in DOFZ silicon detectors

The leakage current of silicon detectors scales with the rate of generation of primary defects, and is roughly independent on particle spectra as could be seen from the examination of Figures 2b and 3b.

In Figures 4 and 5, the time evolution of effective carrier concentrations after irradiation, calculated in the same hypotheses and environments, is presented. For both materials, for rates of generation of defects below $10^5$ pairs/cm$^3$/sec (corresponding to cosmic fields and ILC), the inversion does not occur, and for DOFZ material the modification of $N_{eff}$ is irrelevant.
The predicted results suggest that the DOFZ Si is radiation harder only in conditions of weaker radiation fields (as cosmic and ILC ones) and the behaviour of DOFZ silicon detectors is satisfactory in the first months of operation at $\mu^+\mu^-$ and LHC and its upgrades. For longer time of irradiation at these last facilities, it is not possible to make a choice between FZ and DOFZ materials. So, DOFZ technology is more useful for low rates of primary defects and/or relatively short times of exploitation.

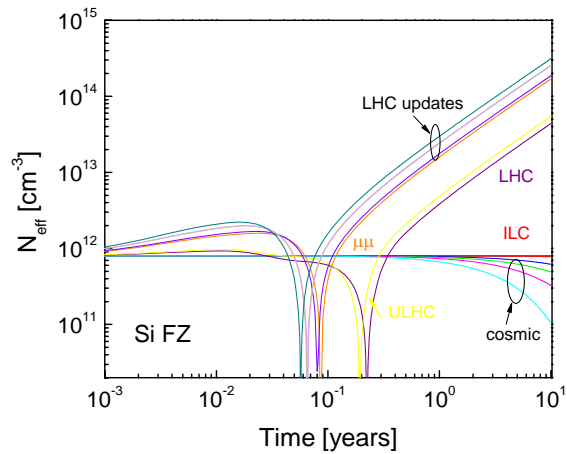

**Figure 4**
Time dependence of $N_{eff}$ in FZ Si detectors during continuous irradiation in different environments

If silicon detectors are directly exposed to primary proton fluxes from cosmic rays, the dominant contribution to degradation comes from protons with low energy, and increases with the distance from the Earth. In the cabin, the rates of generation of defects are minor (comes essentially from secondary particles) and the DOFZ material is certainly more adequate for these applications.

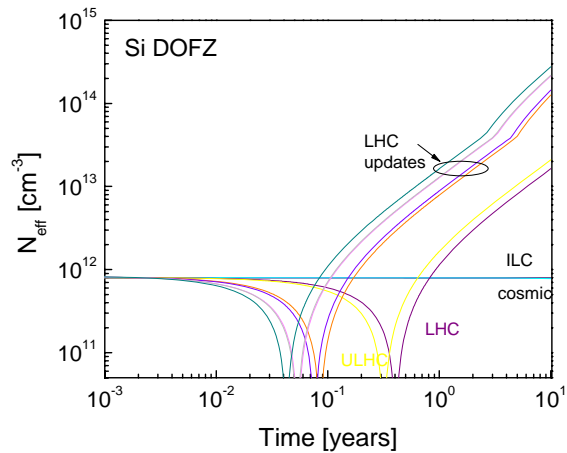

**Figure 5**
Time dependence of $N_{eff}$ in FZ Si detectors during continuous irradiation in different environments

From these calculations we didn't found at the highest rates of generation of defects considered here any clear correlation of the degradation with the initial concentration of oxygen in silicon.

Because these processes have temperature dependence, the decrease of the temperature is strongly required, in order to diminish these macroscopic effects.

## 5. Conclusions and Summary

The main features of microscopic and macroscopic degradation of detectors, after lepton and hadron irradiations, produced in silicon grown by FZ and DOFZ technologies could be reproduced in the frame of the present model. The pre-existence of phosphorus, oxygen and carbon impurities, the generation of primary defects: self-interstitial, classical and fourfold coordinated vacancy - $Si_{FFCD}$ thermally and in irradiation processes, as well as their kinetics are major physical aspects that must be considered.

The effects of the radiation in silicon are dependent on the rate of generation of primary defects, but they are neither clearly correlated with the concentration of oxygen in silicon – except the case of low rates or short time of irradiation, nor with the resistivity of the starting material in the range $1 \div 15$ KΩcm. The identity of the primary particle which initiated the bulk processes is lost after the primary interaction.

For continuous irradiation, the obtained results suggest that the differences between DOFZ and FZ silicon could be observed only in the first months of operation, where DOFZ silicon seems radiation harder. For long times of irradiation and / or high fluences, it is not possible to make a choice from FZ and DOFZ materials. DOFZ technology is more useful for low rates of primary defect generation and in conditions of relatively short times of exploitation.

The calculated degradation of the leakage current of silicon detectors scales with the rate of primary defect generation.

In agreement with model predictions, the contribution of primary defects is important and must be considered.

These processes have temperature dependence and the decrease of the temperature is strongly required in order to diminish these macroscopic effects.